\documentclass{article}
\usepackage{spconf,amsmath,graphicx}
\usepackage{comment}
\usepackage{amsfonts}
\usepackage{threeparttable}

\title{Deep-based quality assessment of medical images through domain adaptation}
\name{Marouane Tliba$^{1}$, Aymen Sekhri$^{2}$, Mohamed Amine KERKOURI $^{1}$,  Aladine Chetouani$^{1}$}
\address{$^{1}$Laboratoire PRISME, Université d'Orléans, Orléans, France\\
$^{2}$Institut National Des Télécommunications et TIC, Oran, Algeria\\
}
\begin{document}
%\ninept
%
\maketitle
\begin{abstract}

Predicting the quality of multimedia content is often needed in different fields. In some applications, quality metrics are crucial with a high impact, and can affect decision making such as diagnosis from medical multimedia. In this paper, we focus on such applications by proposing an efficient and shallow model for predicting the quality of medical images without reference from a small amount of annotated data. Our model is based on convolution self-attention that aims to model complex representation from relevant local characteristics of images, which itself slide over the image to interpolate the global quality score. We also apply domain adaptation learning in  unsupervised and semi-supervised manner. The proposed model is evaluated through a dataset composed of several images and their corresponding subjective scores. The obtained results showed the efficiency of the proposed method, but also, the relevance of the  applying domain adaptation to generalize over different multimedia domains regarding the downstream task of perceptual quality prediction. \footnote{Funded by the TIC-ART project, Regional fund (Region Centre-Val de Loire)}

%Predicting the quality of multimedia content is often needed in different contexts. In some applications, 
%quality metrics are crucial with a high impact.In this paper, we focus on medical images and propose an efficient and shallow model for predicting the quality of such data with reference. Our model is based on a self-attention module that aims to focus on the more relevant characteristics of the images. We also apply domain adaptation in unsupervised and semi-supervised manner. The proposed model is evaluated through a dataset composed of several images and their corresponding subjective scores. The results obtained showed the efficiency of the proposed approach and the relevance of applying domain adaptation.

\end{abstract}
\vspace{-3mm}
\section{Introduction}
\label{sec:intro}
\vspace{-3mm}

Predicting the quality of multimedia content is often needed in several fields \cite{Chetouani20EUSIPCO,chetouani2018ICIP}. It allows to quantify how much the introduced distortions on a given multimedia may damage the visual perceived quality. It is used to enhance the quality of experience \cite{Qualinet}, optimize compression schemes \cite{Suiyi20CompQua} or employ as pre-treatment for some computer vision applications \cite{Chetouani18MEDPRAI} such as biometrics \cite{Fourati19MTAP}, Face recognition \cite{Khodabakhsh19IQAFACES} and so forth. For some specific applications, quality metrics are crucial with a high impact. Medical images are among the more sensitive data since their quality may lead to wrong diagnosis and prognosis \cite{leveque2018subjective}. It is therefore important to develop effective metrics dedicated to the data domain of these particular images.

Depending on the availability of the pristine image, existing metrics can be divided into three main categories: Full Reference metrics where the pristine image is supposed accessible, Reduced Reference metrics where only some features of the pristine image is given and No Reference or blind metrics where only the distorted image is available. Here, we focus on blind approach since its corresponds more to the real case. Interesting metrics were already proposed in the literature. In \cite{Outtas18Dataset}, the authors proposed to extend the well-known NIQE method \cite{niqe} for medical images. The metric, called NIQE-k, is based on a signal analysis in frequency domain. In \cite{kohler2013automatic}, the authors developed a metrics based on a gradient analysis, while texture features were used in \cite{remeseiro2017objective}.

In this paper, we propose an efficient and shallow model for predicting the quality of medical images without reference from small amount of annotated data. Well-known deep convolutional neural network (CNN) architectures such as VGG \cite{VGG} or ResNet \cite{resnet} are designed to work well on hierarchical representation learning. Their initial layers detect simple patterns like edges and gradients, while higher layers detect more abstract features related the global structure \cite{Yosinski2015UnderstandingNN}. The robustness of these models may lead to ignore the introduced perceptual effects. To address this issue in our context, we employ shallow CNN models that incorporate a self-attention module, but also which itself slide over the image in order to model complex local features related to quality prediction downstream task. Nonetheless, lake and shortness of existing medical imaging quality dataset inhibit the development of personalized deep quality metrics. As a solution, we apply domain adaptation learning in unsupervised and semi-supervised manner in order to create a link between the data domains. The proposed model is evaluated through a dataset composed of several images and their corresponding subjective scores. The results obtained showed its efficiency and the relevance of applying domain adaptation to reduce the shift between data domains distribution.
 
 \begin{figure*}[h]
  \centering
  \centerline{\includegraphics[ width = \textwidth,height = 55mm]{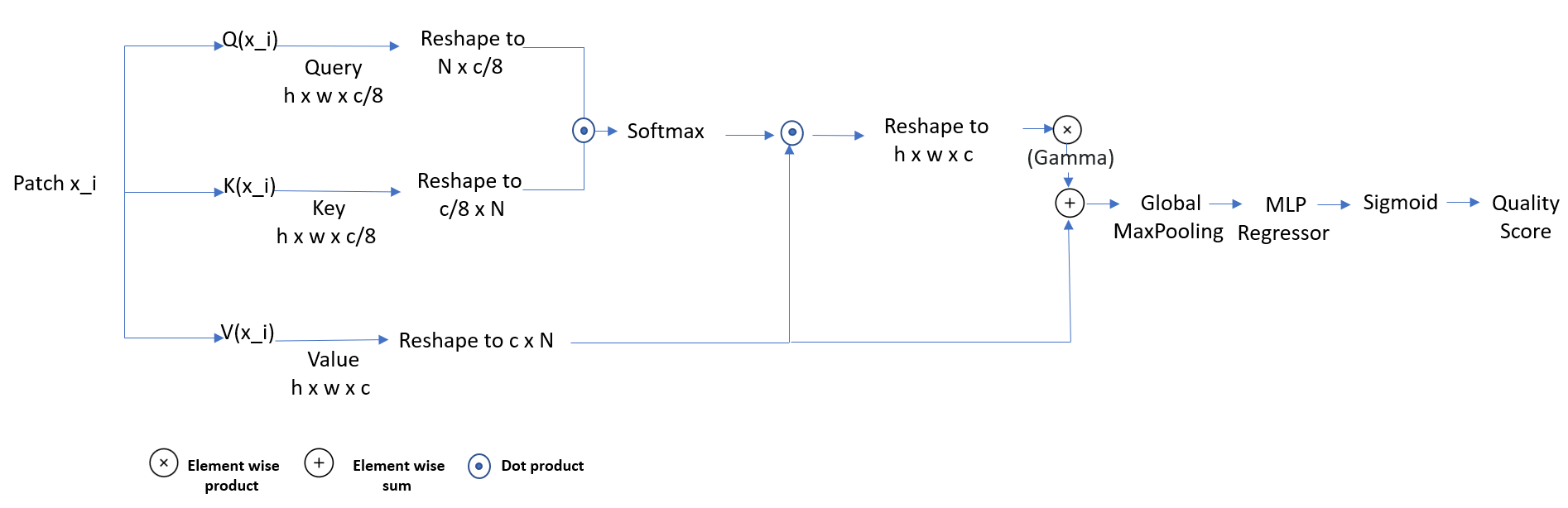}}
\caption{ Model Architecture : Self Attention Quality Metric (SAQM)}
\vspace{-3mm}
\label{fig:arch}

\end{figure*}

The main contributions of our paper are summarized below:\vspace{-3mm}

\begin{itemize}
    \item We propose a novel shallow and efficient CNN model for predicting perceptual quality, relying on extracting useful information from distorted local features.\vspace{-3mm}

    \item We incorporate self-attention module in order to focus on the main intrinsic characteristics of the considered images, and model efficient representation from long range of distant low level features.\vspace{-3mm}

    \item We conduct extensive experiments to demonstrate the effectiveness of our approach, and analyse the impact of unsupervised and semi-supervised domain adaptation in reducing the shift in extracting relevant features distribution between medical and natural scene images.
\end{itemize}
\vspace{-4mm}
\section{Proposed Method}
\label{sec:format}
\vspace{-3mm}

\begin{figure}[ht!]
\centering
\includegraphics[width=0.35\textwidth]{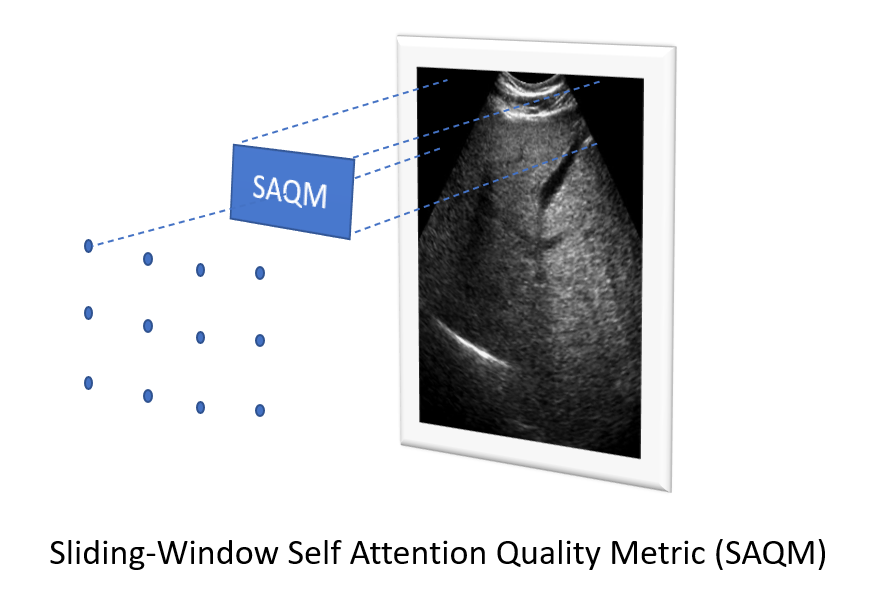}
\caption{Sliding Window Self-Attention Quality Metric}
\label{fig-SlidingWindow}
\end{figure}

The overall architecture of our method is illustrated in Figure \ref{fig:arch}.
The main contribution of our approach is to use a shallow convolution based self-attention module, which itself slide over the image for extraction informative local features, and model the global quality score (see Fig. \ref{fig-SlidingWindow}). Main components of our model are described in details below.

\vspace{-5mm}
\subsection{Deep Quality Self-attention Kernel}

The integration of attention mechanism has recently shown important leap in the performance of various downstream computer vision tasks \cite{Djilali2020ATSalAA} \cite{SALYPATH}. Unlike the absolute attention mechanisms, the one mentioned above learns in a fully adaptive, joint, and task-oriented manner, which allows the network to prioritise and associate weights to feature vectors \cite{satsal}. The self-attention computes the response at a position in a vector or sequence by attending all positions within the same sequence. In greater detail, it draws the relationship between distant features, incorporating self-attention prompts to our shallow network in order to capture complex features related to our downstream task (i.e. quality assessment), thus boosting the representation capability of the full network. 
%The main goal of self-attention is to determine a new set of vector values representing global vector features dependency. Thus, self-attention reveals the set of values to pay more attention to the interaction of input vector features. In simpler words, 

For a given vector, we need to extract Query, Key and Value vectors from it using the shallow CNNs architecture in Table \ref{tab:archi}. The latter measures attention by calculating a similarity between the Query and best related Key features using a score function; The output scores go through the normalisation step to have the sum of probability values equal to one. The final adjusted Value vector is a weighted combination of the previous Value vectors based on the normalised score result \cite{satsal}. % The overall architecture of the proposed extended self-attention is described in figure \ref{fig2}.

Each patch is transformed into three variables, The resulted couple $(Query, Key)$ $\in\mathbb{R}^{{C/8}\times N}$ from Q($X_{i}$) and K($X_{i}$),
respectively, simplifying the dimension of $X_{i}\in\mathbb{R}^{C\times N}$, where N =${(h=8 , w=8)}$ representing the number of feature locations, and {${C}$} the number of output channels from $V(.)$ module. The attention map resulted after normalizing the output of dot product between the $Query$ and $key$ vectors using a Softmax function, where $S$ represents the similarity between the $Query$ and $Key$ feature spaces : 

\begin{equation}
S_{lj} = {Query[{l}]^T. Key[{j}]}
\end{equation}
\begin{equation}
A_{j,l}=\frac{\exp(S_{lj})}{\sum^{N}_{j=1} \exp (S_{lj})},
\end{equation}

The attention map A$\in\mathbb{R}^{N \times N}$ represents the likelihood that a particular positional feature in $l$$_t$$_h$ location appears in the  $j$$_t$$_h$  location in $N$ feature locations, ($_{j,l}$)$\in\mathbb{R}^{N}$. The $Value$ feature space is further enhanced by multiplying it to the attention map. A learnable parameter $\gamma$ is also used in order to learn how much the overall patch prediction should relay on the context composed from local features.   

\begin{equation}
Output = {\gamma \times Value. A+ Value }
\end{equation} 
Finally, a Global Max-pooling is applied to the Output in order to have one dimensional vector, which is later passed through a shallow multi-layer perceptron (MLP) regressor to interpolate the patch quality score, the global quality score is obtained by averaging all patches' quality.
\vspace{-3mm}

\begin{table}[htbp!]
\small
\begin{center}
\begin{tabular}{ l c c  }
\hline
\textbf{Layer} & \textbf{Output Shape   } & \textbf{Param \# }   \\ 
\hline

           Conv2d-1  &      [-1, 128, 32, 32]      &      3,584\\ \hline
              ReLU-2 &      [-1, 128, 32, 32]     &           0\\ \hline
            Conv2d-3 &     [-1, 256, 32, 32] &         295,168\\ \hline
              ReLU-4 &          [-1, 256, 32, 32] &               0\\ \hline
            Conv2d-5  &         [-1, 256, 32, 32] &         590,080\\ \hline
              ReLU-6  &         [-1, 256, 32, 32] &               0\\ \hline
            Conv2d-7  &         [-1, 256, 32, 32] &         590,080\\ \hline
              ReLU-8   &        [-1, 256, 32, 32] &               0\\ \hline
         MaxPool2d-9 &          [-1, 256, 16, 16] &               0\\ \hline
           Conv2d-10  &         [-1, 512, 16, 16] &       1,180,160\\ \hline
             ReLU-11   &        [-1, 512, 16, 16] &               0\\ \hline
           Conv2d-12 &          [-1, 512, 16, 16] &       2,359,808\\ \hline
             ReLU-13  &         [-1, 512, 16, 16] &               0\\ \hline
           Conv2d-14  &         [-1, 512, 16, 16] &       2,359,808\\ \hline
             ReLU-15  &         [-1, 512, 16, 16] &               0\\ \hline
        MaxPool2d-16  &           [-1, 512, 8, 8] &               0\\ \hline
           Conv2d-17   &           [-1, C, 8, 8] &          262,656\\ \hline
             ReLU-18 &             [-1, C, 8, 8] &               0\\ \hline
    \end{tabular}
    \begin{tablenotes}
      \small
      \item Q(x), K(x), and V(x) have similar architecture. For Q(x), K(x) the number of channels (C = 512) is divided by 8, to reduce the dimension during measuring the similarity between the Key and Query vectors.
    \end{tablenotes}

\caption{\label{tab:archi} Architecture of our models.}

\end{center}
\vspace{-6mm}
\end{table}

\vspace{-3mm}
\subsection{Domain adaptation}

%The training of our model on natural scene and medical images independently yields to the results shown in Table 1, which represent the cross data-set validation. However, preliminary test results of the same model on images belonging to other source domains did not perform well, as well as training the model only on shallow medical images dataset. 

To fully utilise the power of our models, we need to adapt them to a new source domain. In the first place, the models need to be able to perceive the two domains (i.e. the original natural scene and the medical images) as being part of the same data distribution $\mathcal{D}$ \cite{DA}. That is achieved by minimising the perceived distance between the two distributions of images $\mathcal{D}_n$ $\mathcal{D}_p$.

%The one below is a slightly newer version
As we train our models on images from both distributions, we add a small branch to the networks which classifies the images as being from $\mathcal{D}_n$ or $\mathcal{D}_p$. 
Furthermore, we add a Gradient Reversal Layer (GRL) on top of this branch, which reverses the sign of the gradient flow during back-propagation. Eq. \ref{eq:GRL_forward} defines forward propagation, while Eq. \ref{eq:GRL_backward} concerns back-propagation. 
%\vspace{-3mm}
\begin{equation}
    GRL(x) = x   
    \label{eq:GRL_forward}
\end{equation}
\begin{equation}
    GRL(\frac{\partial L_d}{\partial x}) = - \frac{\partial L_d}{\partial x}   
    \label{eq:GRL_backward}
    %\vspace{-3mm}
\end{equation}
where $x$ is the input of the layer, and  $\frac{\partial L_d}{\partial x}$ represents the gradient of the domain loss $L_d$ when back-propagating through the network.

%down below a reviewd version 
The reversal of the gradient helps the feature extractor of the network to minimise the distance between domain distributions, thus forcing the feature extractor to disregard the domain-specific features and noises, and emphasise the mutual characteristics of the two domains.
%The reversal of the gradient helps the feature extractor of the network to minimize the distance between domain distributions. This forces the feature extractor to disregard the domain specific features and noises and emphasise on the mutual useful features of the 2 domains. 

This can be also modeled as the union of the 2 considered distributions minus the noise distribution of each of the domains: 
\begin{equation}
     \mathcal{D} = \mathcal{D}_p + \mathcal{D}_n - ( \mathcal{N}_p + \mathcal{N}_n )  
    \label{eq:dists}
    \vspace{-1mm}
\end{equation}
where $\mathcal{D}_n , \mathcal{D}_p$ and $\mathcal{D}$ are defined as before, and $\mathcal{N}_p$ and $\mathcal{N}_n$ are the specific noise distributions of the source domain. 

%during this training we used a mix of 2000 Salicon \cite{salicon} images and 2000 unlabeled paintings images scarped from  the internet. 

\subsection{Technical Details}
We implemented our models in PyTorch and trained them on both data domains with and without domain adaptation each time for 100 epochs. The $K(.)$, $Q(.)$ and $V(.)$ modules were randomly initialized, as well as the regressor and the domain classifier. We normalize the ground truth quality scores (i.e. MOS: Mean Opinion Score) from both data domains to have probability distributions. We employed Binary Cross Entropy (i.e. BCE) in order to minimize the global risk, and the Adam optimiser to train the model. We set the learning rate to $5*10^{-4}$, and  the $\gamma$ parameter was initialized to zero in order to focus on learning the main task.

\section{Experimental results}
\label{sec:pagestyle}

\subsection{Datasets}

\begin{figure}
\centering
\includegraphics[width=0.3\textwidth]{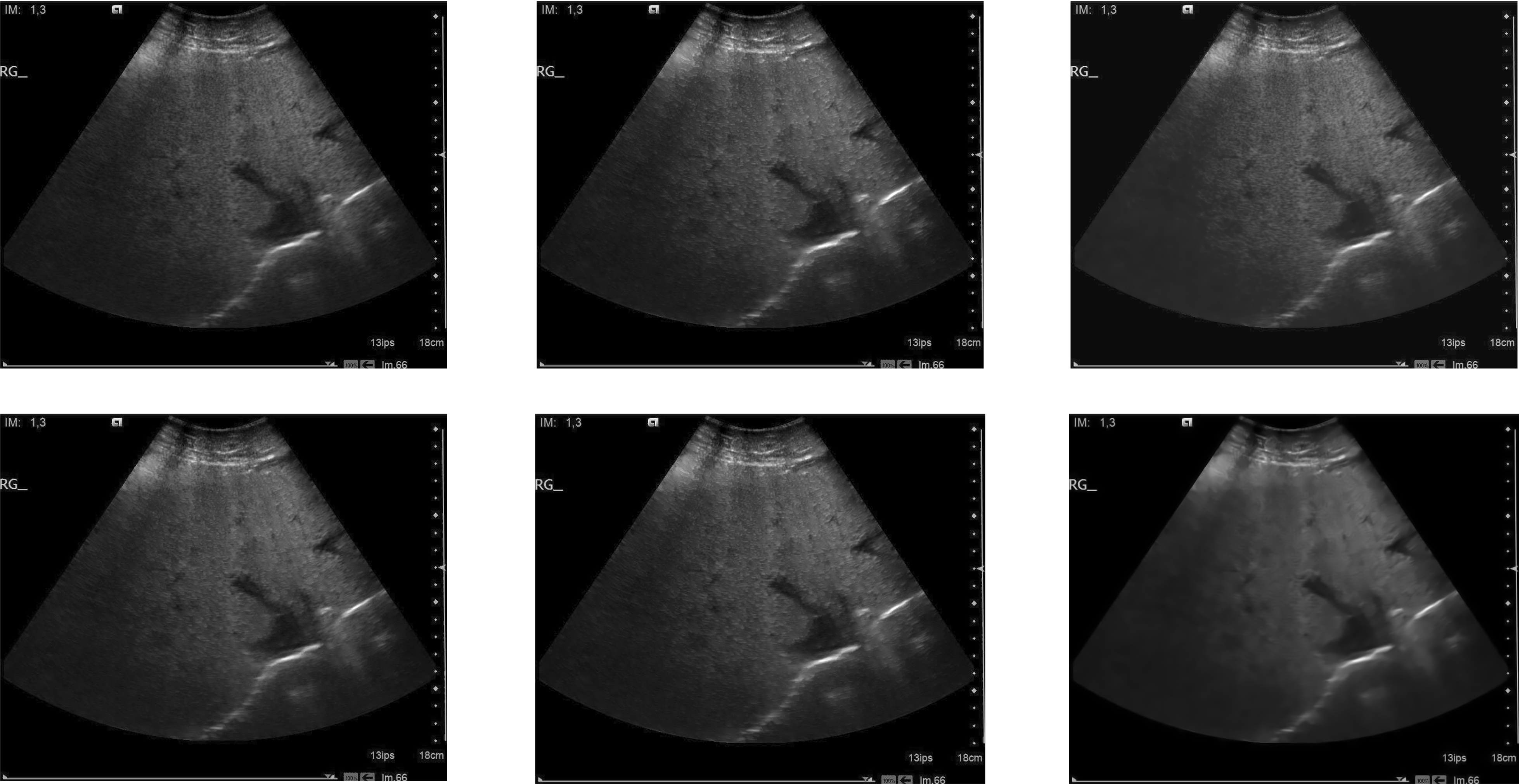}
\caption{Sample of images of the considered dataset.}
\label{fig-sampleMI}
\end{figure}

As mentioned above, domain adaptation has been applied in this study in order to fully optimize the use of our models. To this end, 2 datasets has been used: One composed of natural images considered here as the source data and the second one composed of medical images considered as the target data. Both datasets are briefly described below:
\vspace{-3mm}

\begin{itemize}
    \item \textbf{A subset of TID13 dataset:} Images from TID13 dataset \cite{TID13} has been used in this study. More precisely, we considered only the 125 denoised images (i.e. distortions number 9 of the dataset) derived from 25 reference images. For each image, the corresponding MOS value is given.\vspace{-3mm}

    \item \textbf{MD72 dataset:} The medical dataset proposed in \cite{Outtas18Dataset} has been here used. The latter, denoted here as MD72, is composed of 72 liver Ultrasound images with the corresponding MOS. More precisely, 60 denoised images were derived from 12 pristine images through 5 denoising algorithms. A sample of images is shown in Fig  \ref{fig-sampleMI}. %This dataset was here considered as target dataset for DA learning.
\end{itemize}

It is worth noting that only a specific subset of TID13 dataset has been used as source data for domain adaptation learning. This choice was motivated by the fact that the considered medical image dataset is composed only of denoised images and thus this subset is most related to our downstream task.

In this section, we further analyse the impact of the proposed Domain Adaptation (DA) schemes on the performance. The goal is to show the relevance of using DA in such context. To this end, we computed the correlations for three configurations:
\vspace{-3mm}

\begin{enumerate}
    \item Fully supervised approach \textbf{without DA} by training the model on MD72 and testing it on both datasets.\vspace{-3mm}

    \item Fully supervised approach\textbf{ without DA} by training the model on TID13 and testing it on both datasets.\vspace{-3mm}

    \item \textbf{Unsupervised DA} by using TID13 as source data and MD72 as target data.\vspace{-3mm}
    \item \textbf{Semi-supervised DA} by using TID13 as source data and MD72 as target data.\vspace{-3mm}

\end{enumerate}
\vspace{-1mm}
\begin{table}[htbp!]
\small
\begin{center}
\begin{tabular}{ c c c c c  }
\hline
\textbf{Config.} & \multicolumn{2}{c}{\textbf{MD72}} & \multicolumn{2}{c}{\textbf{TID13}} \\ 
 & \textbf{PLCC  $\uparrow$ } & \textbf{SROCC $\uparrow$ } & \textbf{PLCC  $\uparrow$ } & \textbf{SROCC $\uparrow$ }   \\ 
\hline
 1 &  0.557  &  0.670   & 0.324  &  0.414  \\ \hline
 2 &  0.685  &  0.560   & 0.906   & \textbf{0.794} \\ \hline
 3 & 0.756  &  0.769   & 0.683 & 0.685 \\ \hline
 4 &  \textbf{0.810} &  \textbf{0.812}   & \textbf{0.907}  & 0.784  \\ \hline
\end{tabular}
\caption{\label{tab:Ablation}Impact of the DA on the performance.}
\end{center}
\end{table}
\vspace{-5mm}

From the results shown in Table \ref{tab:Ablation}, several observations can be made. According to the results obtained for configuration 1, we can see that training our model directly on the medical dataset was not enough to predict the quality well, even with a shallow model. The correlations obtained were thus very low for both datasets. The results of configuration 2 show that it seems easier to reach suitable performance for task related to natural images (i.e. subset of TID13 dataset) than medical images. Through the results of configurations 3 and 4, we can clearly see the impact of DA on performance. Indeed, the performance achieved on the target dataset (i.e. MD72) grown significantly. However, we also noticed that the correlations drop on the source data (i.e. the subset of TID13) when the unsupervised learning of DA was applied. The best overall correlations were obtained through semi-supervised DA learning for both datasets with considerable improvements.

\vspace{-3mm}
\subsection{Comparison with state-of-the-art methods}

Our method is evaluated in terms of correlations with the subjective scores only on the target data (i.e. medical images). To this end, we computed the Pearson and Spearman correlations between the predicted scores and subjective ones. We compared the performance of our model to some blind state-of-the-art metrics: BRISQUE \cite{brisque}, NIQE \cite{niqe}, bliinds \cite{bliinds} and BIQAA \cite{BIQAA}. We also considered a metric, so-called NIQEK \cite{Outtas18Dataset}, that was developed specifically for such kind of medical images. In addition to those methods, we finally considered 2 metrics dedicated to the blur (i.e. MARZILIANO \cite{Marziliano} and RadialIndexD \cite{radialindex}) and 1 other for estimating the noise (i.e. \cite{noise}).

The table \ref{tab:SOTAComp} shows the performances obtained for each of the compared methods. The 2 best results are highlighted in bold. As can be seen, all the compared metrics obtained correlations lower than 0.7, except our method which outperformed them by reaching correlations higher than 0.8. The second best PLCC and SROCC were obtained respectively by BIQAA and RadialIndexD, far from the results obtained by our method.

\begin{table}[htbp!]
\small
\begin{center}
\begin{tabular}{ l c c  }
\hline
\textbf{Metric} & \textbf{PLCC  $\uparrow$ } & \textbf{SROCC $\uparrow$ }   \\ 
\hline
 BRISQUE  & 0.6551 &  0.3829 \\ \hline
 NIQE  & 0.4972 &  0.2768 \\ \hline
 NIQEK  & 0.5682 &  0.4429 \\ \hline
 bliinds  &  0.4373 &   0.4131 \\ \hline
 BIQAA  & \textbf{0.6643} &  0.5297 \\ \hline
 Noise  & 0.5978 &  0.5286 \\ \hline
 MARZILIANOmetricD  & 0.5303 &  0.3028 \\ \hline
 RadialIndexD  & 0.5905 &  \textbf{0.5338} \\ \hline
 Our  & \textbf{0.810}  &  \textbf{0.812}  \\ \hline
\end{tabular}
\caption{\label{tab:SOTAComp} Results obtained on MD72 dataset}
\end{center}
\vspace{-6mm}
\end{table}

\section{Conclusion}
\vspace{-3mm}

In this paper, we proposed a novel method for quality assessment of medical images. In particular, we looked to address the task of predicting the satisfied ultrasound image quality for applying an accurate diagnosis from small amount of annotated data. To this end, we employed a shallow CNNs as a self attention modules in order to model efficiently complex representation from local features that affect the visual perception. We also used unsupervised and semi-supervised domain adaptation learning to reduce the shift between data domain distributions, and generalize well over small medical annotated datasets. The obtained results shows the effectiveness and the efficiency of our method. 
We look forward to create a stronger link between data domains, thought urging our shallow model to learn first the quality relevant common structures of different domains from images only in self-supervised way.  
\vspace{-3mm}
%\newpage
\label{sec:typestyle}

\label{sssec:subsubhead}

\bibliographystyle{IEEEbib}
\bibliography{strings,refs}

\end{document}